\documentstyle[12pt,epsf,epic,curves]{article}
\def\1ad{\mbox{\normalsize $^1$}}
\def\2ad{\mbox{\normalsize $^2$}}
\def\3ad{\mbox{\normalsize $^3$}}
\def\4ad{\mbox{\normalsize $^4$}}
\def\5ad{\mbox{\normalsize $^5$}}
\def\6ad{\mbox{\normalsize $^6$}}
\def\7ad{\mbox{\normalsize $^7$}}
\def\8ad{\mbox{\normalsize $^8$}}
\def\makefront{\vspace*{1cm}\begin{center}
\def\newtitleline{\\ \vskip 5pt}
{\Large\bf\titleline}\\
\vskip 1truecm
{\large\bf\authors}\\
\vskip 5truemm
\addresses
\end{center}
\vskip 1truecm}
\newcommand{\grapha}{
\setlength{\unitlength}{1.2mm}
\begin{picture}(10,4)
\put(2.5,0.5){\bigcircle{5}}
\put(7.5,0.5){\bigcircle{5}}
\end{picture}
}

\newcommand{\frgrapha}{
\setlength{\unitlength}{1.2mm}
\begin{picture}(10,4)
\curvedashes[0.8mm]{0,1,1.7}
\put(2.5,0.5){\bigcircle{5}}
\put(7.5,0.5){\bigcircle{5}}
\end{picture}
}

\newcommand{\graphb}{
\setlength{\unitlength}{1.3mm}
\begin{picture}(12,5)
\put(2,0.5){\bigcircle{4}}
\drawline(4,0.5)(8,0.5)
\drawline(4,0.5)(4,-3.5)
\drawline(3.5,-3.0)(4.5,-4.0)
\drawline(3.5,-4.0)(4.5,-3.0)
\put(10,0.5){\bigcircle{4}}
\drawline(8,0.5)(8,-3.5)
\drawline(7.5,-3.0)(8.5,-4.0)
\drawline(7.5,-4.0)(8.5,-3.0)
\end{picture}
}

\newcommand{\frgraphb}{
\setlength{\unitlength}{1.3mm}
\begin{picture}(12,3)
\curvedashes[0.8mm]{0,1,1.7}
\put(2,0.5){\bigcircle{4}}
\dashline[50]{0.8}(4,0.5)(8,0.5)
\put(10,0.5){\bigcircle{4}}
\end{picture}
}

\newcommand{\graphc}{
\setlength{\unitlength}{1.2mm}
\begin{picture}(10,5)
\curve(0,0.5,1,2.6,2,3.3,3,3.7,4,3.9,5,4,6,3.9,7,3.7,8,3.3,9,2.6,10,0.5)
\drawline(0,0.5)(10,0.5)
\curve(0,0.5,1,-1.6,2,-2.3,3,-2.7,4,-2.9,5,-3.0,6,-2.9,7,-2.7,8,-2.3,9,
        -1.6,10,0.5)
\drawline(0,0.5)(0,-4.5)
\drawline(-0.5,-4.0)(0.5,-5.0)
\drawline(-0.5,-5.0)(0.5,-4.0)
\drawline(10,0.5)(10,-4.5)
\drawline(9.5,-4.0)(10.5,-5.0)
\drawline(9.5,-5.0)(10.5,-4.0)
\end{picture}
}

\newcommand{\frgraphc}{
\setlength{\unitlength}{1.2mm}
\begin{picture}(10,5)
\curvedashes[0.8mm]{0,1,1.7}
\curve(0,0.5,1,2.9,2,3.7,3,4.2,4,4.4,5,4.5,6,4.4,7,4.2,8,3.7,9,2.9,10,0.5)
\dashline[50]{0.8}(0,0.5)(10,0.5)
\curve(0,0.5,1,-1.9,2,-2.7,3,-3.2,4,-3.4,5,-3.5,6,-3.4,7,-3.2,8,-2.7,9,-1.9,10,0.5)
\dashline[50]{0.8}(0,0.5)(10,0.5)
\end{picture}
}

\newcommand{\graphd}{
\setlength{\unitlength}{1.4mm}
\begin{picture}(12,7)
\put(1,0.5){\bigcircle{2}}
\put(1,0.5){\circle*{1}}
\drawline(2,0.5)(6,0.5)
\put(8,0.5){\bigcircle{4}}
\drawline(6,0.5)(6,-3.5)
\drawline(5.5,-3.0)(6.5,-4.0)
\drawline(5.5,-4.0)(6.5,-3.0)
\end{picture}
}
\begin {document}
\def\titleline{Critical Phenomena, Strings, and Interfaces}
\def\authors{Gernot M\"unster}
\def\addresses{
Institut f\"ur Theoretische Physik I,
Universit\"at M\"unster\\
Wilhelm-Klemm-Str.~9, D-48149 M\"unster, Germany
}
%
%
\makefront
%
\section{Introduction}
Traditionally, the participants of the ``Ahrenshoop International
Symposium on the Theory of Elementary Particles'' form two nearly
disjoint subsets, consisting of string theorists and lattice gauge
theorists.
So, for a plenary speaker the question arises: is it possible to give a
talk which addresses both of these species?
I don't have an answer, but this contribution is meant as an attempt.

Considering the basic theoretical objects which are being studied, there
is no apparent relation.
The geometrical objects of string theory are world-sheets of open or
closed strings. We shall not speak about the additional internal degrees
of freedom here. A parameterized world-sheet is described by functions
$x^{\mu}(\sigma,\tau)$.
In lattice gauge theory, on the other hand, the basic objects are group
valued variables $U(x,\mu)$ associated with the links $(x,\mu)$ of a
space-time lattice. That looks very different.
\begin{figure}[h,t]
\centering
\epsfxsize 12cm
\epsffile{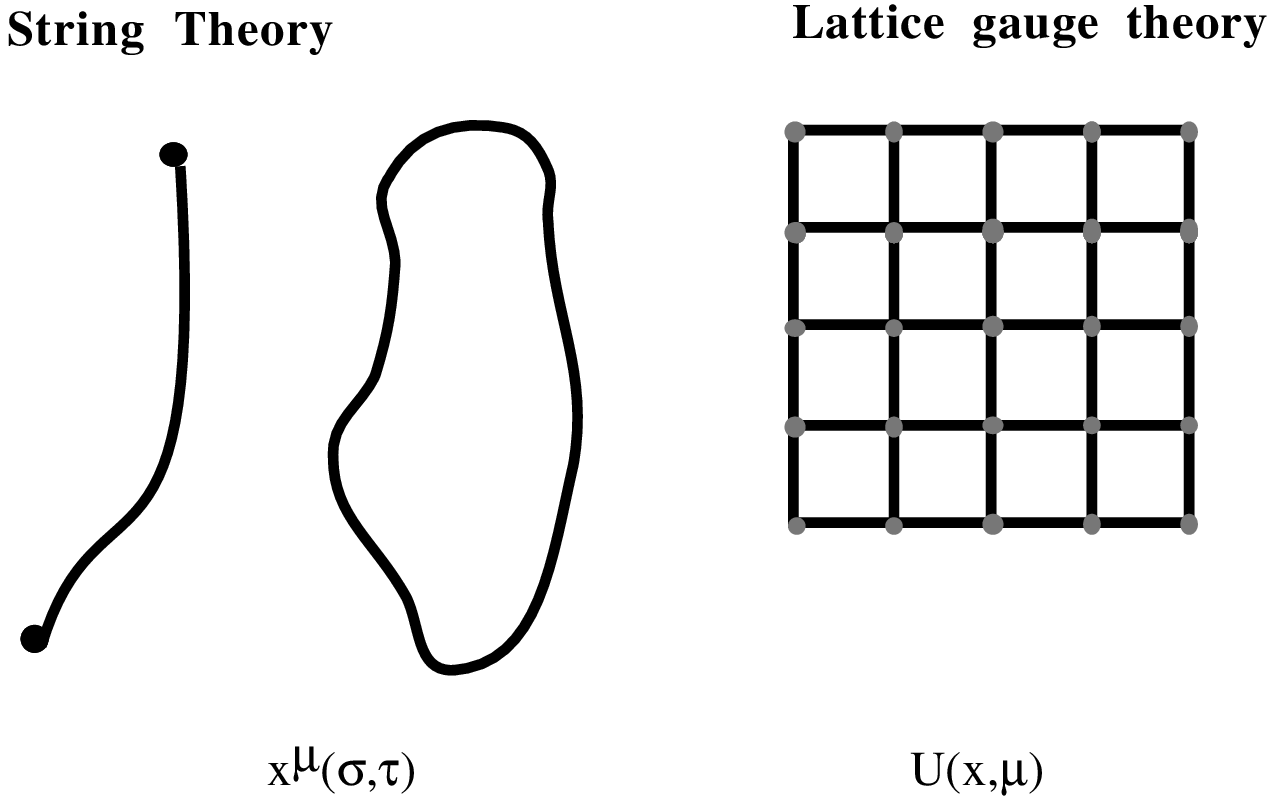}
\end{figure}

Let us consider bosonic string theory in $d$ dimensional space-time a
little bit closer. The Nambu-Goto action of a
world sheet, parameterized by $x^{\mu}(\sigma,\tau)$, with $\mu$ running
from 1 to $d$, is
\begin{equation}
S = \alpha \cdot \mbox{Area} = \alpha \int\!\!d\sigma d\tau \sqrt{g}\,,
\end{equation}
where
\begin{equation}
g = \det\left( \frac{\partial x_{\mu}}{\partial \xi_{\alpha}}
               \frac{\partial x^{\mu}}{\partial \xi_{\beta}} \right),
\hspace{10mm}
\xi_{\alpha} = \sigma, \tau .
\end{equation}
Most people would start from the Polyakov action \cite{Polyakov}
nowadays, but let us stick to the Nambu-Goto action for the time being.

To quantize string theory basically means to give meaning to
functional integrals of the type
\begin{equation}
\label{Zstring}
Z = \int\!\!D[x^{\mu}(\sigma,\tau)]\ e^{-S} \,.
\end{equation}
As is known since long there are obstacles to naive quantization for
any dimension $d$.
After employing reparametrization invariance to fix the so-called
conformal gauge, the remaining conformal symmetry is generated by
operators $L_n$, which obey the Virasoro algebra
\begin{equation}
[L_n , L_m] = (n-m) L_{n+m} + \frac{d-26}{12} (m^3-m) \delta_{n+m,0}\,,
\end{equation}
where the ghost contribution to the $L_n$'s is included.
Consistent straightforward quantization requires the central extension
to vanish.
Therefore only
\begin{equation}
d = 26
\end{equation}
is allowed.

Now let us turn to lattice gauge theory.
The link variables $U(x,\mu)$ represent the gauge field $A_{\mu}(x)$ in
the sense of elementary parallel transporters on the lattice:
\begin{equation}
U(x,\mu) \simeq e^{- a A_{\mu}(x)} \in {\rm SU}(N)\,,
\end{equation}
where $a$ is the lattice constant.
The simplest action for a lattice gauge field is the Wilson action
\begin{equation}
S = - \frac{\beta}{N} \sum_{p} {\rm Re\,Tr}\,U(p)\,,
\end{equation}
where $p$ are the elementary plaquettes on the lattice, $U(p)$ is the
ordered product of the four link variables belonging to the boundary of
plaquette $p$, and $\beta$ is inversely proportional to the coupling
constant squared.
The basic functional integral is of the type
\begin{equation}
Z = \int\!\!D[U(x,\mu)] \, e^{-S}\,.
\end{equation}
One method to evaluate such integrals approximatively is the strong
coupling expansion, i.e.\ an expansion in powers of $\beta$, analogous
to high temperature expansions in statistical mechanics.
For example, the strong coupling expansion for Wilson loops leads to
diagrams, which are (more or less) surfaces on the lattice bounded by
the loop. These surfaces look like world-sheets of strings.
Indeed, it turns out that the strong coupling expansion leads to
confinement of static quarks, and the above-mentioned surfaces are
related to confining strings between colour sources \cite{Wilson,KS}.
\begin{figure}[h,t]
\centering
\epsfxsize 10cm
\epsffile{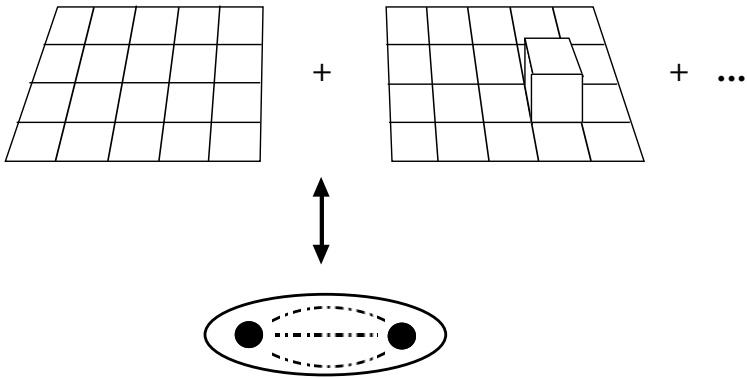}
\parbox[t]{0.9\textwidth}{
\caption{Strong coupling diagrams for a Wilson loop and the confining
string}
}
\end{figure}

So there appears to be some relation between lattice gauge theory and
strings.
Can this relation be made more precise? In particular, is it possible to
describe lattice gauge theory in terms of an effective string theory?
Many attempts have been made in this direction, e.g.\ by Nielsen and
Olesen, 't Hooft, Nambu, Gervais and Neveu, Polyakov, Migdal and others
\cite{strings}. It appears to be difficult to obtain concrete
results.

As mentioned above, strings appear in the strong coupling expansion. It
is, however, difficult to reformulate lattice gauge on the basis of the
strong coupling expansion as a theory of strings. The main problem comes
from the nontrivial weights of the diagrams.

Another point, where strings appear in lattice gauge theory, is the
$1/N$-expansion \cite{dWtH}, but it is difficult to obtain a string
formulation for finite $N$ from that.

My impression is that the question of an effective string theory for
gauge fields is still open.
Let us therefore turn to simpler field theoretic models and look for
strings in them. In particular, let us consider scalar fields.
The simplest model with a scalar field is the Ising model. Its field
$s(x)$ is associated with the points of a lattice and only assumes
values $s(x) = \pm 1$, representing spins pointing up or down.
The action is given by
\begin{equation}
S = - \kappa \sum_{<xy>} s(x) s(y)\,,
\end{equation}
where $<xy>$ is a link between nearest neighbour points $x$ and $y$ on
the lattice.

In the same universality class is $\phi^4$-theory. In the phase with
broken symmetry the action can be written as
\begin{equation}
\label{action}
S = \int\!\!d^{d}x \left\{ \frac{1}{2} (\partial_{\mu} \phi)^2
+ \frac{g}{4!} (\phi^2 - v^2)^2 \right\}\,.
\end{equation}
The minima of the double well potential are located at $\phi = \pm v$.

In the Ising model at large $\kappa$ (``low temperatures''), as well as
in $\phi^4$-theory in the broken symmetric phase interfaces appear.
They are $(d-1)$-dimensional surfaces separating regions with opposite
values of the field. In the Ising model they are domain walls between
regions with $s(x) = +1$ and $s(x) = -1$. For large enough
$\kappa$ fluctuations are small and these interfaces are rather well
defined objects. On a finite rectangular lattice with appropriate
boundary conditions nearly flat interfaces can be prepared.
Similarly, in the $\phi^4$-model interfaces separate regions with
$\phi(x) \approx +v$ from those with $\phi(x) \approx -v$.
\begin{figure}[h,t]
\centering
\epsfxsize 10cm
\epsffile{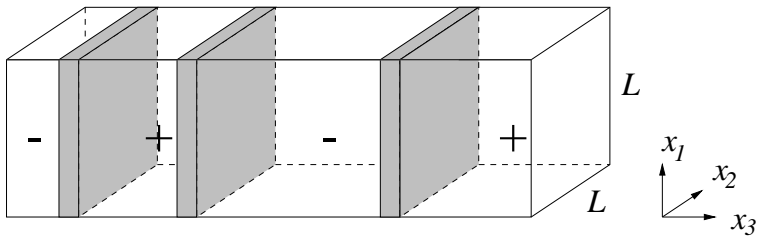}
\parbox[t]{0.8\textwidth}{
\caption{Interfaces in a finite Ising-like system}
}
\end{figure}

The interfaces of scalar field theory are string-like, i.e.\
two-dimensional, only for $d=3$, whereas the strings of lattice gauge
theory are always two-dimensional.  In $d=3$, where we have both types
of ``strings'', there is an interesting additional relation between
them.
\begin{figure}[h,t]
\centering
\epsfxsize 10cm
\epsffile{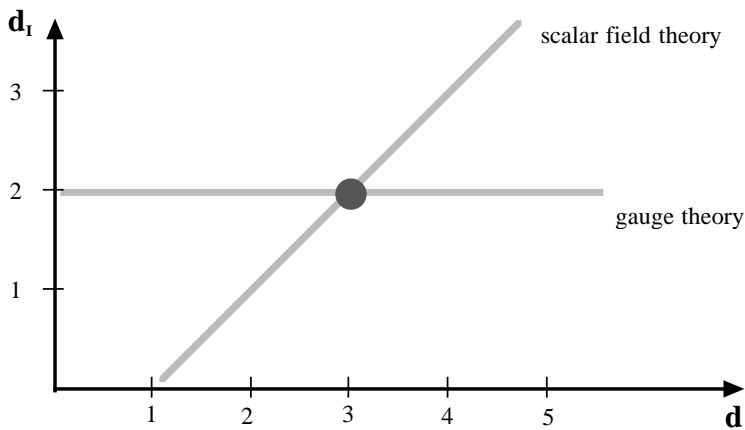}
\parbox[t]{0.8\textwidth}{
\caption{Dimensionality $d_I$ of surfaces in $d$ dimensions}
}
\end{figure}
This is duality, which can be made quite explicit for models with
an abelian symmetry group.  An abelian model, whose fields are p-forms
and whose interaction terms are defined on $({\rm p}+1)$-cells, is
mapped by duality onto an equivalent model of $(d-{\rm p}-2)$-forms and
interactions on $(d-{\rm p}-1)$-cells.  For $d=3$ duality maps the Ising
model (p=0) at low temperatures onto the $Z_2$ gauge theory at high
temperatures and vice versa.  Therefore the Ising string in 3 dimensions
is really the same as the $Z_2$ gauge theory string.

In the following we shall concentrate on the three-dimensional situation
more closely.
%
\section{Interfaces in $d=3$}
There are many systems of statistical mechanics in three dimensions
for which interfaces play an interesting role.
These include
\begin{itemize}
\item liquid gas coexistence
\item binary liquid mixtures
\item anisotropic ferromagnetism
\item ferroelectrics
\item superconductors
\item crystal growth.
\end{itemize}

Near a critical point ($T \approx T_c$) of such a system one observes
universal behaviour of certain quantities related to interfaces.
Consider the interface tension $\tau$ or the reduced interface tension
$\sigma = \tau /k T$, where $k$ is Boltzmann's constant.
It is positive for $T < T_c$, but vanishes according to
\begin{equation}
\sigma \sim \sigma_0 \left| \frac{T-T_c}{T_c} \right|^{\mu}
\end{equation}
as the temperature $T$ approaches $T_c$.
The value of the critical exponent is approximately $\mu \approx 1.26$
and appears to be universal. On the other hand, the amplitude $\sigma_0$
is not universal.

The critical law for the correlation length $\xi$, which diverges at the
critical point, is
\begin{equation}
\xi \sim \xi_0 \left| \frac{T-T_c}{T_c} \right|^{-\nu}\,.
\end{equation}
Widom's scaling law \cite{Widom} relates the indices $\mu$ and $\nu$. In
$d=3$ it reads
\begin{equation}
\mu = 2 \nu\,.
\end{equation}
Consequently the product $\sigma \xi^2$ approaches a finite value at the
critical point:
\begin{equation}
\sigma \xi^2 \longrightarrow \sigma_0 \xi_0^2 \doteq R_-\,.
\end{equation}
The number $R_-$ appears to be universal, too. It is a so-called
universal amplitude product.
In the past it has been studied experimentally for various systems, by
means of Monte Carlo calculations, and by field theoretic methods.
Other quantities, which have been investigated in connection with
interfaces, are the interface width, the interface profile, the
interface stiffness etc..
Closely related to $R_-$ is
\begin{equation}
R_+ = \sigma_0 (\xi_0^+)^2 =
R_- \left( \frac{\xi_0^+}{\xi_0^-}\right)^2\,,
\end{equation}
where $\xi_0^+$ is the correlation length amplitude of the high
temperature phase (which is easier accessible experimentally), and
$\xi_0^-$ the one of the low temperature phase.

Why should one study $R_-$ or $R_+$?
Early results from the $\epsilon$-expansion \cite{BF} and from
Monte-Carlo calculations \cite{Binder} were in strong conflict with
experimental numbers $R_+ = 0.38(2)$ (for a brief summary of the history
and relevant references see \cite{Jue}).
Therefore the question of universality for these interface-related
quantities arose.
Furthermore it became desirable to learn about the status of the
theoretical predictions.

So, let us turn to the theoretical calculation of the interface tension.
%
\section{Description of interfaces by field theory}
The critical phenomena of systems in the universality class of the
three-dimensional Ising model, like those mentioned in the previous
section, can be calculated in the framework of massive $\phi^4$-theory.
The scalar field $\phi(x)$ represents the order parameter. For example
in the case of a liquid mixture this would be the difference of
densities of the two liquids.
The action for the scalar field is given in Eq.\,(\ref{action}).
In order to study a planar interface we consider the theory in a
rectangular box with quadratic cross-section $L^2$ in the $x_1 -
x_2$-plane and antiperiodic boundary conditions in the
$x_3$-direction.
Alternatively, one might choose
\begin{equation}
\phi(x) \longrightarrow \left\{
\begin{array}{l}
+v, \hspace{4mm} \mbox{as}\ x_3 \rightarrow \infty\\
-v, \hspace{4mm} \mbox{as}\ x_3 \rightarrow -\infty\,.
\end{array} \right.
\end{equation}
A classical solution of the field equations is given by the kink
\begin{equation}
\phi_c = \sqrt{\frac{3 m^2}{g}} \tanh \frac{m}{2} (x_3 - a)\,,
\end{equation}
centered at $x_3 = a$, where $m^2=g v^2 / 3$.
\begin{figure}[h,t]
\centering
\epsfxsize 10cm
\epsffile{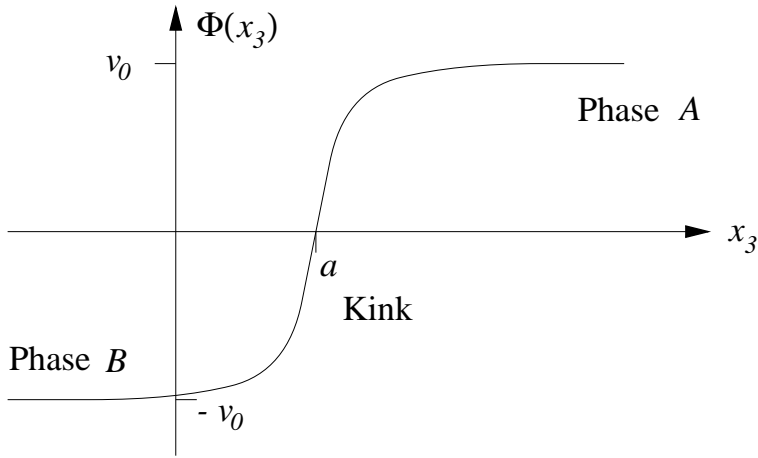}
\parbox[t]{0.8\textwidth}{
\caption{The kink as an interface between two phases}
}
\end{figure}

Its classical action is
\begin{equation}
S_c = 2 \frac{m^3}{g} L^2\,.
\end{equation}
Thus the saddle point approximation to the functional integral with the
boundary conditions specified above,
\begin{equation}
Z_{+-} = \int_{\pm} D\phi\ e^{-S}\,,
\end{equation}
is given by
\begin{equation}
Z_{+-} \approx e^{-S_c}\,.
\end{equation}
Because $Z_{+-}$ is the partition function of a system with an interface,
it should depend on $L$ like
\begin{equation}
Z_{+-} \sim e^{-\sigma L^2}\,.
\end{equation}
We can read off the interface tension in the saddle point approximation
\begin{equation}
\sigma \approx 2 \frac{m^3}{g}\,,
\end{equation}
and for the product of interest we obtain
\begin{equation}
\sigma \xi^2 = \frac{\sigma}{m^2} \approx 2 \frac{m}{g}\,.
\end{equation}
Introducing the dimensionless coupling constant
\begin{equation}
u \equiv \frac{g}{m}\,,
\end{equation}
we write our tree level result as
\begin{equation}
\frac{\sigma}{m^2} \approx \frac{2}{u}\,.
\end{equation}

Now let us take into account fluctuations
\begin{equation}
\phi(x) = \phi_c(x) + \eta(x)
\end{equation}
around the classical solution.
The action
\begin{equation}
S = S_c + \frac{1}{2} \int\!\!d^3 x \, \eta(x) M \eta(x) + O(\eta^3)
\end{equation}
contains the fluctuation operator
\begin{equation}
\label{opM}
M = - \partial_{\mu} \partial^{\mu} + m^2
- \frac{3}{2} m^2 \cosh^{-2}(\frac{m}{2} x_3)\,.
\end{equation}
For the case of periodic boundary conditions, where an interface need
not be present, the partition function is denoted $Z_{++}$, and $M$ is
replaced by the Helmholtz operator
\begin{equation}
M_0 = - \partial_{\mu} \partial^{\mu} + m^2\,.
\end{equation}
The relevant ratio of partition functions can then expressed as
\begin{eqnarray}
\frac{Z_{+-}}{Z_{++}} &=& \sqrt{\frac{S_c}{2\pi}}
\left( \frac{\det' M}{\det M_0} \right)^{-1/2}
\times \exp \Bigg\{ - S_c
+\;\frac{1}{2}\;\graphd\;+\;
        \frac{1}{8}\;\bigg[\;\grapha\;-\;\frgrapha\;\bigg]
\nonumber\\[1.5ex]
&& +\;\frac{1}{8}\;\bigg[\;\graphb\;-\;\frgraphb\;\bigg]
    \;+\;\frac{1}{12}\;\bigg[\;\graphc\;-\;\frgraphc\;\bigg]
 + O(g^2) \Bigg\} ,
\end{eqnarray}
where the meaning of the graphs can be found in \cite{HM}.
To evaluate this expression, first of all renormalization has to be
carried out in the usual way.
Moreover, the contribution of multi-kink configurations has been taken
into account, but I refuse to reveal any details here.

Luckily, in the one-loop approximation the determinants can be
calculated exactly \cite{Mue3d} and one obtains
\begin{equation}
\frac{Z_{+-}}{Z_{++}} = C e^{-\sigma L^2}\,,
\end{equation}
with
\begin{equation}
\frac{\sigma}{m^2} = \frac{2}{u_R} \left( 1 + \sigma_1 \frac{u_R}{4\pi}
+ \ldots \right)
\end{equation}
and exact expressions for the constants $C$ and $\sigma_1$. The
quantities $m$ and $u_R$ are the renormalized mass and coupling now.

Believe it or not, it has been possible to evaluate the two-loop
contribution to the ratio $\sigma/m^2$, too \cite{HM}.
Written in the form
\begin{equation}
\sigma_2 \left( \frac{u_R}{4\pi} \right)^2
\end{equation}
the coefficient $\sigma_2 = - 0.0076(8)$ turns out to be rather small.
In order to obtain the desired universal ratio we have to evaluate the
function
\begin{equation}
\frac{\sigma}{m^2} \equiv R_-(u_R)
\end{equation}
at the fixed point value $u_R = u_R^*$. The most recent value
$u_R^* = 14.3(1)$ from Monte Carlo calculations \cite{Caselle} is
consistent with an estimate of 14.2(2) from three-dimensional field
theory \cite{GKM}.
At this value of the coupling the one-loop contribution is 24\% and the
two-loop contribution roughly 1\% of the tree-level term.
The final result from field theory for the amplitude product is
\begin{equation}
R_- = 0.1065(9)\,.
\end{equation}
It compares well with the recent Monte Carlo result $R_- = 0.1040(8)$
by Hasenbusch and Pinn \cite{HP1}.
The corresponding numbers for $R_+$, using theoretical values for
$\xi_0^+ / \xi_0^-$, lie in the range from 0.40(1) to 0.42(1) and are
compatible with the recent experimental result $R_+ = 0.41(4)$
\cite{Mainzer}, which is higher than the earlier average of 0.37(3).
%
\section{Effective string description}
Field theory describes fluctuating interfaces, as we have seen.
The relevant partition function is analogous to a functional integral
over fluctuating string world-sheets, Eq.\,(\ref{Zstring}).
A proposal to describe the dynamics of fluctuating interfaces in terms
of an effective string model is the ``capillary wave model'' or
``drumhead model'' \cite{BLS}.
The interface is considered to be a surface without overhangs, which can
be described by a height function $x_3 = h(x_1,x_2)$.
\begin{figure}[h,t]
\centering
\epsfxsize 10cm
\epsffile{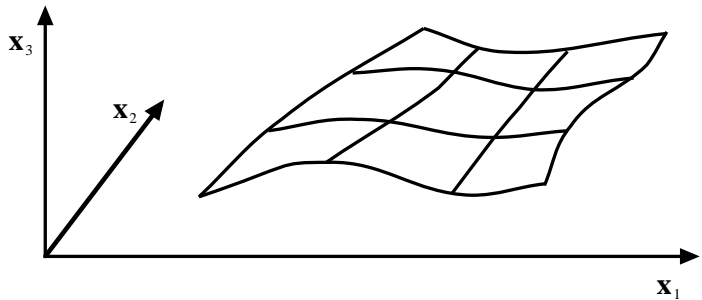}
\parbox[t]{0.8\textwidth}{
\caption{Smooth surface in the drumhead model}
}
\end{figure}
The action is, as in the Nambu-Goto case, given by the area:
\begin{equation}
S = \tilde{\sigma} \int_0^{L_1}\!\!dx_1 \int_0^{L_2}\!\!dx_2
\sqrt{1 + \left( \frac{\partial h}{\partial x_1} \right)^2
        + \left( \frac{\partial h}{\partial x_2} \right)^2 }\,.
\end{equation}
Expanding in powers of $h$ we get
\begin{eqnarray}
S &=& \tilde{\sigma} L_1 L_2 +
\frac{\tilde{\sigma}}{2}\int\!\!dx_1 \int\!\!dx_2
\left[ \left( \frac{\partial h}{\partial x_1} \right)^2
     + \left( \frac{\partial h}{\partial x_2} \right)^2 \right]
\nonumber\\
&& - \frac{\tilde{\sigma}}{8}\int\!\!dx_1 \int\!\!dx_2
\left[ \left( \frac{\partial h}{\partial x_1} \right)^2
     + \left( \frac{\partial h}{\partial x_2} \right)^2 \right]^2
+ \cdots
\end{eqnarray}
The second term, the Gaussian action $S_G$, is quadratic in the field
$h(x_1,x_2)$ and describes a massless scalar field in two dimensions.
The expansion of the action above leads to an expansion of the partition
function which is organized in powers of $1/\tilde{\sigma} L_1 L_2$:
\begin{equation}
Z_{+-} = e^{- \tilde{\sigma} L_1 L_2}\ Z_1 \cdot Z_2 \cdot \ldots
\end{equation}
where the one-loop term
\begin{equation}
Z_1 = \int\!\!Dh\, e^{-S_G} = (\det \Delta_{(2)})^{-1/2}
\end{equation}
can be expressed in terms of the determinant of the
two-dimensional Laplace operator with appropriate boundary conditions.
This is a well known object in 2d conformal invariant field theory with
central charge $c=1$ on a torus \cite{IZ}, and has been calculated
explicitly:
\begin{equation}
Z_1 =
\frac{1}{\sqrt{-i \tau}} \left| \frac{\eta(\tau)}{\eta(i)} \right|^{-2}
\cdot \mbox{const.}\,,
\end{equation}
where
\begin{equation}
\eta(\tau) = q^{1/24} \prod_{n=1}^{\infty} (1 - q^n)\,, \hspace{5mm}
q=e^{2\pi i \tau}
\end{equation}
is Dedekind's eta-function, and the parameter $\tau$ is given by the
aspect ratio
\begin{equation}
\tau = i \frac{L_1}{L_2}\,.
\end{equation}
This opens the possibility to test the capillary wave model by studying
the dependence of $Z_1$ on $L_1 / L_2$.
Comparing two partition functions with equal area, $L^2 = L_1 L_2$, the
leading term cancels out and one has
\begin{equation}
\frac{Z_{+-}(L_1,L_2)}{Z_{+-}(L,L)} = \sqrt{\frac{L_2}{L_1}}
\left| \frac{\eta(i L_1 / L_2)}{\eta(i)} \right|^{-2}\,.
\end{equation}
A comparison of this formula with Monte Carlo results for the $d=3$
Ising model shows very good agreement, supporting the capillary wave
model \cite{CGV}.

The capillary wave model also predicts the roughening phenomenon
\cite{BLS}. The width $w$ of an interface of size $L \times L$, given
by
\begin{equation}
w^2 = \frac{1}{L^2} \int\!\!dx_1 dx_2\
\langle (h(x_1,x_2) - \langle h \rangle )^2 \rangle \,,
\end{equation}
can be calculated in the Gaussian capillary wave model \cite{LMW,HP2}.
For large $L$ it diverges like
\begin{equation}
w^2 = \frac{1}{2 \pi \sigma} \log \frac{L}{R_0}\,,
\end{equation}
with some cutoff length $R_0$.
This behaviour indicates the dominance of longwave fluctuations of the
interface.

The Gaussian approximation, considered so far, is not specific to the
capillary wave model.
In fact, it is an infrared fixed point for a whole class of effective
models.
In order to test the capillary wave model one should go beyond the
Gaussian approximation.
This has been done in \cite{CFGHPV} (see also \cite{DF}).
They have evaluated the two-loop contribution $Z_2$ to the partition
function and get
\begin{eqnarray}
Z_2 &=& 1
+ \frac{\tilde{\sigma}}{8}\int\!\!dx_1 \int\!\!dx_2\
\langle ((\nabla h)^2)^2 \rangle_G \nonumber\\
&=& 1 + \frac{1}{4 \sigma L^2}
\end{eqnarray}
for $L_1 =L_2 = L$.
Included is a renormalization of the interface tension $\sigma$.
This term has been nicely confirmed by Monte Carlo calculations
\cite{CFGHPV}.

At this point one might wonder whether it is possible to derive
the string description, i.e.\ the capillary wave model, directly from
$\phi^4$-theory in three dimensions.
In the Gaussian approximation this can indeed be done, see \cite{PV}.
The fluctuation operator $M$, Eq.\,(\ref{opM}), decomposes into the
two-dimensional Laplacean $\Delta_{(2)}$ and a Schr\"odinger operator
$Q$: $M=-\Delta_{(2)}+Q$.
The operator $Q$ has two discrete eigenvalues, 0 and $3 m^2/4$, and a
continuous spectrum. The zero-mode is associated with translations of
the interface along $x_3$.
Consequently, the determinant of $M$ contains $\det \Delta_{(2)}$
as a factor. But this is just the contribution of a two-dimensional free
massless scalar field, and is identical to the Gaussian capillary wave
model.
The remaining factor belongs to massive modes on the interface and does
not dominate the long-wavelength behaviour. It contributes to the
renormalization of $\sigma$.

To derive the string model from $\phi^4$ theory beyond the Gaussian
approximation of interface fluctuations is more difficult.
First of all, for a given field $\phi(x)$ the interface variables
$h(x_1,x_2) = F[\phi(x)]$ have to be defined suitably, at least for
fields not too far away from the kink solution $\phi_c$.
Formally one would then write
\begin{equation}
\int\!\!D[\phi(x)]\ e^{-S} \equiv \int\!\!D[h(\xi)]\ e^{-S_h}
\end{equation}
with
\begin{equation}
e^{-S_h} =
\int\!\!D[\phi(x)]\ e^{-S} \delta[h(\xi) - F[\phi(x)]\,.
\end{equation}
This approach has been studied e.g.\ in \cite{Diehl,Zia}.
In the low temperature limit, $T \rightarrow 0$, the interface (or
string) action $S_h$ indeed approaches the Nambu-Goto action
\begin{equation}
S_h = \sigma \cdot \mbox{Area} + \mbox{corrections}\,.
\end{equation}
This is, however, far away from the critical point.
%
\section{Questions}
Many questions concerning the relation of critical interfaces to strings
are still open.
Let us consider fluctuations. Near the critical point the fluctuations
of the interface are strong.
In the field theoretic approach this means that a typical field $\phi$
is by no means similar to the classical solution $\phi_c$.
Why then, does the semiclassical expansion work so well?

We can interpret this effect as a result of renormalization.
Fluctuations on short scales produce ultraviolet divergencies, which
lead to the renormalization of the mass $m$ and the interface tension
$\sigma$.
In the renormalized propagator in a kink background the UV-fluctuations
are summed up effectively.
In terms of renormalized quantities we can thus expect a smoother
behaviour. So it is the usual picture of renormalization, which is at
work.

On the side of the string description the same question arises. Near the
critical point the interface is far from smooth. Overhangs, bubbles and
handles appear.
Why does the capillary wave model work?
\begin{figure}[h,t]
\centering
\epsfxsize 10cm
\epsffile{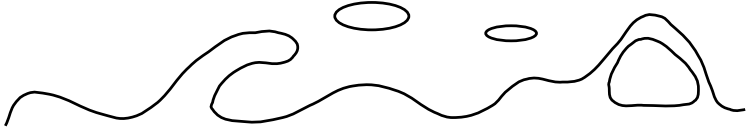}
\parbox[t]{0.8\textwidth}{
\caption{Overhangs, bubbles and handles}
}
\end{figure}

Attempts to answer this question have been made in \cite{handles}.
The analogue of renormalization is claimed to be a ``condensation of
handles''.
Near each point of the interface thin tubes can be attached, which
represent overhangs or handles.
They yield a renormalization factor proportional to the area, which in
turn can be absorbed into the renormalization of the tension $\sigma$.
The average size of a handle is expected to be microscopic and
independent of the large scale geometry of the interface, so that these
condensed handles do not influence the long-wavelength behaviour.

Another question concerns the conformal anomaly.
The bosonic string can be quantized consistently without Liouville modes
only in 26 dimensions.
Why does the effective string model work in 3 dimensions?

A more careful treatment of the transformation from field variables to
string variables would take into account the arising Jacobian $J$:
\begin{equation}
e^{-S_h} =
\int\!\!D[\phi(x)]\ e^{-S} J[h(\xi)]\,\delta[h(\xi) - F[\phi(x)]]\,.
\end{equation}
A calculation of $J$ in the framework of a four-dimensional abelian
Higgs-model \cite{Akhmedov} gave
\begin{equation}
J = \mbox{const.} \exp \left\{ \int\!\!d^2 \xi \frac{22}{96 \pi}
(\partial_a \log \sqrt{g} )^2 + \ldots \right\} .
\end{equation}
This term contributes to the Virasoro generators $L_n$, and their
algebra reads
\begin{equation}
[L_n , L_m] = (n-m) L_{n+m} +
\frac{d-26+22}{12} (m^3-m) \delta_{n+m,0}\,.
\end{equation}
The anomaly term now vanishes in $d=4$ dimensions, as desired.
The string action is given by
\begin{equation}
S_h = \int\!\!d^2 \xi \left\{ \sigma \sqrt{g}
- \frac{11}{48 \pi} (\partial_a \log \sqrt{g} )^2
- \beta \sqrt{g} (\partial_a t_{\mu \nu})^2 + \ldots \right\}\,,
\end{equation}
where $t_{\mu \nu}$ is the extrinsic curvature.
The second piece, the Liouville term, has been first proposed in
\cite{PS}.

Because the Jacobian $J$ is of a pure geometric nature, the same effect
is expected to take place in three-dimensional $\phi^4$-theory.

To summarize, there are interesting relations between string theory and
fluctuating interfaces in critical statistical systems, and there are
several open points, which deserve further study.

%
\end{document}